# THE APM CLUSTER SURVEY: CLUSTER DETECTION AND LARGE-SCALE STRUCTURE

G. B. Dalton
*Departme... ord, UK.*

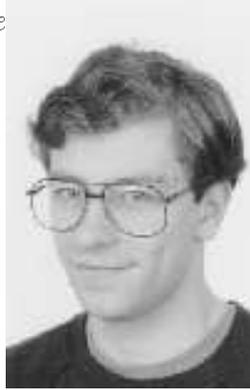

**Abstract**
The APM Cluster Survey was based on a modification of Abell's original classification scheme for galaxy clusters. Here we discuss the results of an investigation of the stability of the statistical properties of the cluster catalogue to changes in the selection parameters. For a poor choice of selection parameters we find clear indications of line-of-sight clusters, but there is a wide range of input parameters for which the statistical properties of the catalogue are stable. We conclude that clusters selected in this way are indeed useful as tracers of large-scale structure.

## 1 Introduction

For many years, all statistical analyses of the distribution of clusters of galaxies has relied on the work of [1] and [2]. In recent years, however, the uniformity of these catalogues, and hence their usefulness as tools for studying the large-scale structure, have come into question ([7], [10] and [6]). With the aim of resolving these questions and producing a uniform cluster catalogue we have applied a modified version of Abell's algorithm to the APM Galaxy Survey ([8] and [9]). Here we describe an investigation of the properties of a wide range of possible catalogues based on different parameter choices. We find the statistical properties of the catalogue to be fairly stable to changes of the input parameters, but we will also describe some potential pitfalls. A full account of this work will be published in [5].

## 2 The Cluster Selection Algorithm

Abell's algorithm defined clusters in terms of the number of galaxies seen on the photographic plate above the estimated background, within two magnitudes of $m_3^\star$, and within a projected radius of 1.5h$^{-1}$Mpc at the distance estimated from the magnitude of $m_{10}$. For a range of projected radii, $r_C$, we estimate the cluster distance from $m_X$, where $X$ is a function of the cluster richness ($X = \mathcal{R}/\kappa$), which is in turn determined from a magnitude range about $m_X$. This definintion is recursive, and so we apply the convergence criteria that $\delta m_X < 0.025$ mag and that the centroid of the galaxies within the projected radius changes by $< 40''$.

---

$^\star$We use the notation $m_i$ to denote the $i$th ranked galaxy in the cluster field after subtraction of the background.

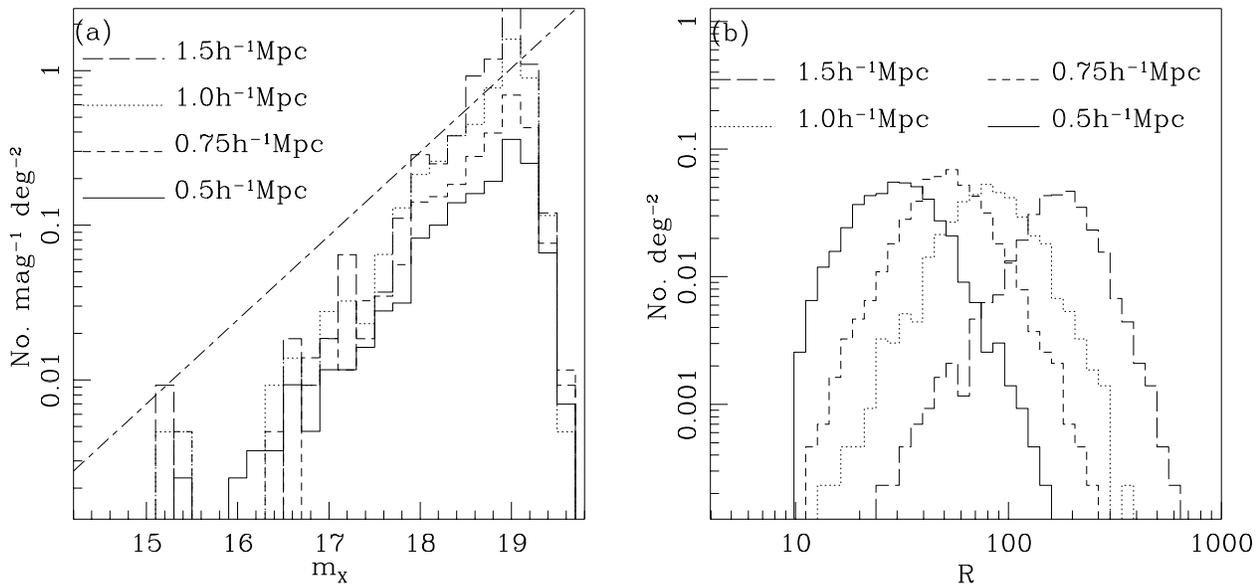

Figure 1: **(a)** The number-magnitude histograms for catalogues with different search radii (see legend), $X = \mathcal{R}/4$. and richnesses determined from the range $[m_X - 0.5, m_X + 1.5]$. The magnitude limits adopted for comparison samples are 19.0, 18.9, 18.8, and 18.7, respectively. The diagonal shows the Euclidean slope of 0.6 for comparison, and successive histograms have been offset by 0.3 in $\log N$ for clarity. For each catlogue we only show data for clusters above the richness limit adopted for the comparison. **(b)** The number-richness histograms for these catalogues, with magnitude limits as above. The richness limits for 500 clusters to these magnitude limits are 190, 99, 67, and 39, respectively.

## 3  Comparison Samples and Completeness

For each catalogue we calibrate the distance estimator as described by [4]. We expect a catalogue to be useable out to the point at which the faint limit of the richness determination is equal to the magnitude limit of the APM Galaxy Survey ($b_J = 20.5$). We compare the properties of groups of catalogues by selecting equal numbers of clusters from each catalogue to a common estimated distance. Figure 1 shows the number-magnitude and number-richness relations for four catalogues with different selection radii (see figure caption). For each catalogue the $N(m_X)$ relation rises smoothly to $m_X = 19.0$ where the catalogue definition becomes invalid. The $N(\mathcal{R})$ relations show that each catalogue is still complete in terms of richness at the comparison limit.

The limiting feature of the catalogues shown in Figure 1 is that the richness is defined by a broad magnitude range about $m_X$ which hits the survey magnitude limit by a redshift of $z \sim 0.1$. We attempted to improve on this limit by reducing the magnitude range by 0.5mag at the faint end to give a limiting $m_X$ of 19.5. However, since the definition of $m_X$ is tied to the richness count, it then becomes necessary to adopt a different value for $X$ to prevent $m_X$ from sampling the luminosity function at an absolute magnitude brighter than $M_*$. If this is not corrected, the cluster selection algorithm becomes extremely sensitive to spurious, line-of-sight associations of galaxies in preference to real clusters [3]. This effect is most noticeable for a large selection radius. Reducing $X$ to a value closer to $X = 2$ compensates adequately for the change in the richness definition, but careful tuning is required to optimise the desired gain in the catalogue depth. One should also be aware that the depth of the catalogue is increased at the expense of increased noise in the richness estimates for nearby clusters, as the background count in a small magnitude range becomes uncertain at brighter magnitudes.

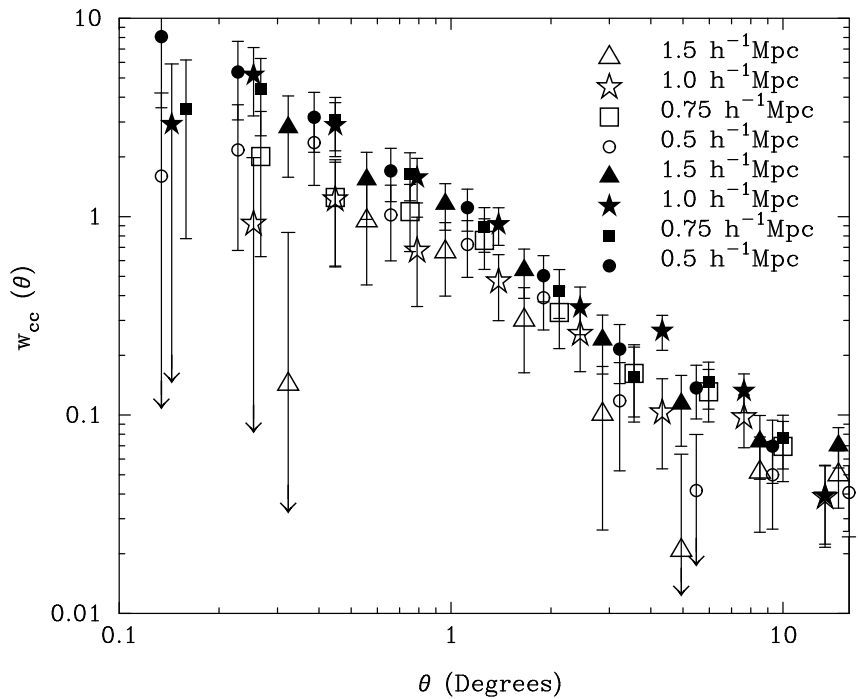

Figure 2: The angular correlation functions for the comparison samples shown in Figure 1 (solid symbols). Also shown are the data for similar catalogues generated with richnesses estimated from the range $[m_X - 0.5, m_X + 1.0]$ (open symbols). Note the depression of clustering due to spurious line-of-sight associations.

## 4 Angular Clustering Measures

For each comparison set of catalogues we calculated the two-point angular correlation functions, $w(\theta)$, using the standard, $(DD/RR)-1$, estimator. Figure 2 shows the results for the catalogues shown in Figure 1, together with results for similar catlogues generated using the smaller magnitude range for the richness estimator as discussed above. It can be seen that the original catalogues (filled symbols) are in good agreement with one another over the whole range of scales for which estimation is possible, and we may conclude that there is no intrinsic variation in the clustering properties introduced by changing the selction radius. For the second set of catalogues (open symbols) there a systematically lower amplitude for $w(\theta)$ and a much greater variation between the catalogues. As noted above, the largest excursions from the solid symbols occur for those catalogues with large selection radii.

At smaller selection radius, $r_C = 0.75h^{-1}$Mpc, we also investigated the effect of varying $\kappa$.

At the depth of the comparison samples considered for $2 \leq \kappa \leq 4$ we find no variation within the amplitude of $w(\theta)$ within the errors, and conclude that the catalogues are stable to such variation at this selection radius. This implies that the reduced amplitude seen in Figure 2 for small radii catalogues (open circles and squares relative to filled circles and squares) must be generated by incompleteness at the low richness end in these catalogues. This is not surprising, given that in Figure 2 we have matched to the lowest effective depth estimate from Figure 1 and hence we are sampling only those clusters in the deeper catalogue with the noisiest richness estimates.

We conducted similar tests on the angular cluster–galaxy correlation functions. These provide a much clearer signal of the line-of-sight effects discussed above. For the original set of catalogues we see evidence for a shoulder in $w_{cg}(\theta)$ which corresponds to the projected cluster selection radius at the mean depth of the sample considered. This effect is significantly more pronounced for catalogues with large $r_C$, and is almost undetectable by $0.75 h^{-1}$Mpc. For the narrower richness definition we see a similar trend as we increase $r_C$ until we reach $r_C = 1.5 h^{-1}$Mpc. For this catalogue the line-of-sight effects are so pronounced that there is *no signal* in the cross-correlation function beyond this radius.

## 5  Conclusions

We have shown that it is possible to select clusters of galaxies in a way that does not affect the large-scale statistical properties of the distribution. There is some gain to be made by a careful choice of distance estimator, but one must be wary of the potential problems associated with a distance estimator which samples the luminosity function at very bright magnitudes. We conclude that clusters of galaxies selected from a high quality galaxy catalogue may be usefully used as a tracer of the matter distribution on large scales.

## References

[1] Abell G. O., 1958.ApJS, 3, 211.

[2] Abell G. O., Corwin H. C., Olowin R. P., 1989.ApJS, 70, 1.

[3] Dalton G. B., 1992.*D.Phil. thesis*, Oxford University.

[4] Dalton G. B., Efstathiou G., Maddox S. J., Sutherland W. J., 1994.MNRAS, 269, 151.

[5] Dalton G. B., Maddox S. J., Sutherland W. J., Efstathiou G., 1995,in preparation.

[6] Dekel A., Blumenthal G. R., Primack J. R., Olivier S., 1989.ApJ, 338, L5.

[7] Lucey J. R., 1983.MNRAS, 204, 33.

[8] Maddox S. J., Sutherland W. J., Efstathiou G., Loveday J., 1990a.MNRAS, 243, 692.

[9] Maddox S. J., Efstathiou G., Sutherland W. J., 1990.MNRAS, 246, 433.

[10] Sutherland W. J., 1988.MNRAS, 234, 159.